# Long-term Operation of the Multi-Wire-Proportional-Chambers of the LHCb Muon System

F. P. Albicocco[a], L. Anderlini[b], M. Anelli[a], F. Archilli[c], G. Auriemma[d,e], W. Baldini[f,1], G. Bencivenni[a], N. Bondar[g], B. Bochin[g], D. Brundu[h], S. Cadeddu[h], P. Campana[a], G. Carboni[i,j], A. Cardini[h], M. Carletti[a], L. Casu[h], A. Chubykin[g], P. Ciambrone[a], E. Dané[a], P. De Simone[a], M. Fontana[k], P. Fresch[d], M. Gatta[a], G. Gavrilov[g], S. Gets[g], G. Graziani[b], A. Kashchuk[g], M. Korolev[m], S.Kotriakhova[g], E. Kuznetsova[g], A. Lai[h], O. Levitskaya[g], A. Loi[h], O. Maev[g,k,1], D. Maysuzenko[g], G. Martellotti[d], S. Nasybulin[g], P. Neustroev[g], R.G.C. Oldeman[h,n], M. Palutan[a], G. Passaleva[b], G. Penso[d], D. Pinci[d], R. Santacesaria[d], M. Santimaria[a], E. Santovetti[i,j], B. Saitta[h,n], A. Saputi[a], A. Sarti[a,l], C. Satriano[d,e], A. Satta[i], B. Schmidt[k], T. Schneider[k], B. Sciascia[a], A. Sciubba[d,l], R. Vazquez-Gomez[k], S. Vecchi[f], A. Vorobyev[g]

a *Laboratori Nazionali dell'INFN di Frascati, Frascati, Italy*
b *Sezione INFN di Firenze, Firenze, Italy*
c *Nikhef National Institute for Subatomics Physics, Amsterdam, Netherlands*
d *Sezione INFN di Roma La Sapienza, Roma, Italy*
e *Università della Basilicata, Potenza, Italy*
f *Sezione INFN di Ferrara, Ferrara, Italy*
g *Petersburg Nuclear Physics Institute, NRC, Kurchatov Institute (PNPI, NRC, KI), Gatchina, Russia*
h *Sezione INFN di Cagliari, Cagliari, Italy*
i *Sezione INFN di Roma Tor Vergata, Roma, Italy*
j *Università di Roma Tor Vergata, Roma, Italy*
k *European Organization for Nuclear Research (CERN), Geneva, Switzerland*
l *Università di Roma la Sapienza, Roma, Italy*
m *Institute of Nuclear Physics, Moscow State University (SINP MSU), Moscow, Russia*
n *Università di Cagliari, Cagliari, Italy*

E-mail: *Oleg.Maev@cern.ch, baldini@fe.infn.it*

ABSTRACT: *The muon detector of LHCb, which comprises 1368 multi-wire-proportional-chambers (MWPC) for a total area of 435 $m^2$, is the largest instrument of its kind exposed to such a high-radiation environment. In nine years of operation, from 2010 until 2018, we did not observe appreciable signs of ageing of the detector in terms of reduced performance. However, during such a long period, many chamber gas gaps suffered from HV trips. Most of the trips were due to Malter-like effects, characterised by the appearance of local self-sustained high currents, presumably originating from impurities induced during chamber production. Very effective, though long, recovery procedures were implemented with a HV training of the gaps in situ while taking data. The training allowed most of the affected chambers to be returned to their full functionality and the muon detector efficiency to be kept close to 100%. The possibility of making the recovery faster and even more effective by adding a small percentage of oxygen in the gas mixture has been studied and successfully tested.*

---

[1] Corresponding authors

## Contents

## 1. Introduction

A primary goal of the LHCb experiment is to look for indirect evidence of new physics in CP violation and rare decays of beauty and charm hadrons. This requires the LHCb detectors to operate in a very challenging radiation environment, as a results of their forward coverage, with a pseudo-rapidity range from 2 to 5. Many of the important physics channels are identified by their very clean muon signatures, so the performance of the muon system is extremely important.

With its 1368 multi-wires-proportional-chambers (MWPC), the muon detector of LHCb [1] is one of the largest instruments of its kind worldwide, and one of the most irradiated. For most of the LHC data taking period we recorded data at an instantaneous luminosity of $4 \times 10^{32}$ cm$^{-2}$ s$^{-1}$. The most irradiated MWPCs integrated ~ 0.6 C/cm of charge per unit length of wire over the past nine years. In this period the MWPC chambers did not show a gain reduction or other apparent symptoms of reduced performance in terms of efficiency or time resolution. The kind of tests performed are described in references [2, 3].

However, during such a long period of running, many gas gaps were affected by the sudden appearance of high currents. This effect, originating from localized areas in the individual gaps of the MWPCs, results in an increased noise rate and a trip of the HV supply system due to a current exceeding the set threshold. The observed phenomenon of high currents that are triggered by high levels of radiation, and that are self-sustained even when the particle flux is reduced when the proton beam had disappeared, suggests that most of the trips are due to Malter-like effects (ME) [4, 5]. The Malter effect is often associated to ageing but in our case there are numerous indications that prolonged irradiation is not the underlying cause.

During operation of the LHC, about 100 gas gaps per year were affected by HV trips. If not treated appropriately, this would have led to a visible loss in detector efficiency. Most of the problematic chambers could be recovered successfully in situ during data taking under nominal beam conditions, by means of a long HV training performed on the affected MWPC gaps. As will be discussed in this paper, this method has proven to be very effective, allowing recovery of normal operating conditions for most of the MWPC gaps affected by HV trips, so that the muon detection efficiency could be kept close to 100%. This is a remarkable result, since the recovery of gas discharge detectors without disassembling is essential for continuous operation of modern experiments.

An overview of the LHCb muon detector is reported in section 2. This is followed by a detailed discussion on the chamber training procedures and the statistics of HV trips and recoveries, in sections 3 and 4. Finally, section 5 reports the results of a test with an accelerated recovery procedure based on the addition of a small fraction of oxygen to the gas mixture. The benefit of using $CF_4$ in the gas mixture to suppress Malter currents and the effect of adding oxygen in the mixture are explained in appendices A and B, respectively.

## 2. The LHCb muon detector

The muon detector of the LHCb experiment [1, 2] consists of five stations M1−M5, placed along the beam axis. Station M1 is located in front of the calorimeters and is used to improve the measurement of the muon transverse momentum for the first level trigger. Stations M2−M5 are placed behind the hadronic calorimeter and are interleaved with iron absorbers to select penetrating muons. As shown in figure 1, each station is divided into four regions R1−R4. The area of these four regions scales, from R1 to R4, with the ratios 1:4:16:64, while the irradiation per unit area decreases. In total, the muon detector is equipped with 20 types of chambers, varying mainly in size. MWPCs are used everywhere, with the exception of region R1 of station M1, where triple-GEMs [6] were adopted due to the higher irradiation.

The whole detector comprises 1380 chambers (1368 MWPCs and 12 GEMs), for a total active area of $435 m^2$. Since the requirements on spatial resolution and rate capability, as well as constructional constraints, vary a lot in different stations and regions, different chamber segmentations and different readout techniques were employed. In the outer regions (R4) there are large wire pads while in the inner regions there are small cathode pads or even smaller logical pads obtained by crossing cathode pads with narrow wire strips [1]. The number of MWPCs installed in each region of the detector, their relevant characteristics and the charge integrated so far per unit length of wire, are listed in table 1.

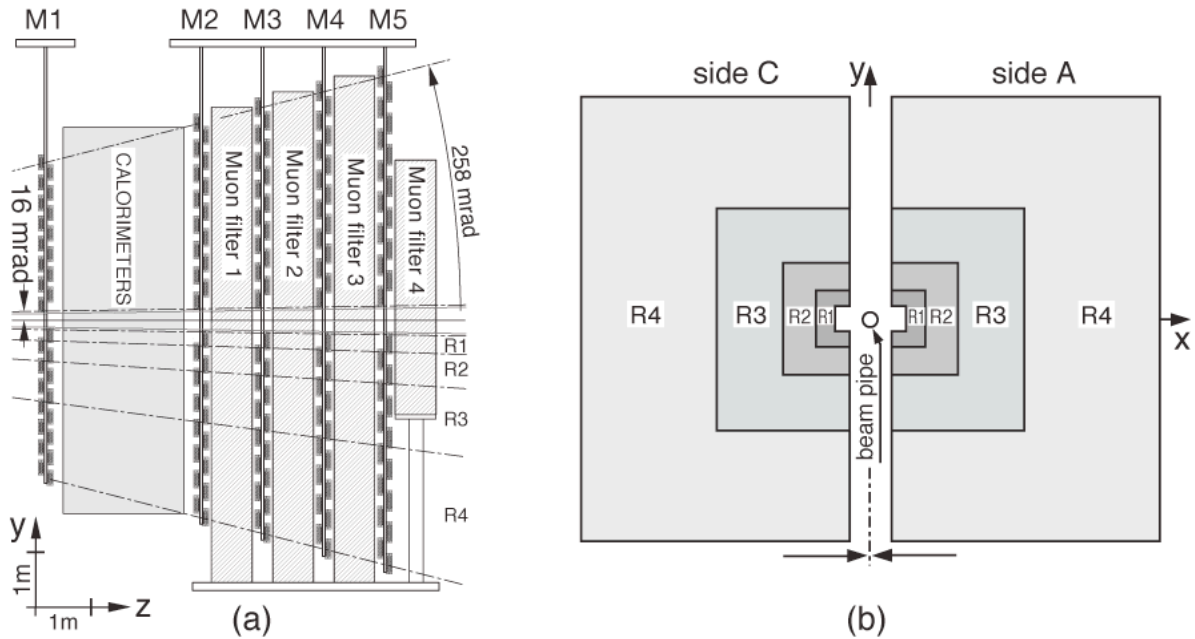

**Figure 1.** (a) Side view of the LHCb muon detector. (b) Station layout with the four regions R1−R4 indicated.

**Table 1.** Relevant parameters for all of the 19 types of MWPCs in the LHCb muon detector. In the last column the charge integrated during the full data taking is reported, ranging from the least to the most irradiated chamber of each region.

| Station | Region | # Chambers | Chamber active area (cm$^2$) | # Gaps per chamber | Readout type | $Q_{int}$ (mC/cm) |
|---|---|---|---|---|---|---|
| M1 | R2 | 24 | 48×20 | 2 | cathode | 95 − 630 |
| M1 | R3 | 48 | 96×20 | 2 | cathode | 20 − 220 |
| M1 | R4 | 192 | 96×20 | 2 | wire | 4 − 80 |
| M2 | R1 | 12 | 30×25 | 4 | mixed | 56 − 150 |
| M2 | R2 | 24 | 60×25 | 4 | mixed | 18 − 80 |
| M2 | R3 | 48 | 120×25 | 4 | cathode | 1.3 − 14 |
| M2 | R4 | 192 | 120×25 | 4 | wire | 0.15 − 1.6 |
| M3 | R1 | 12 | 32×27 | 4 | mixed | 13 − 50 |
| M3 | R2 | 24 | 65×27 | 4 | mixed | 3 − 22 |
| M3 | R3 | 48 | 130×27 | 4 | cathode | 0.13 − 2.2 |
| M3 | R4 | 192 | 130×27 | 4 | wire | 0.03 − 0.3 |
| M4 | R1 | 12 | 35×29 | 4 | cathode | 9 − 40 |
| M4 | R2 | 24 | 70×29 | 4 | cathode | 1 − 10 |
| M4 | R3 | 48 | 139×29 | 4 | cathode | 0.1 − 1.6 |
| M4 | R4 | 192 | 139×29 | 4 | wire | 0.02 − 0.2 |
| M5 | R1 | 12 | 37×31 | 4 | cathode | 12 − 34 |
| M5 | R2 | 24 | 74×31 | 4 | cathode | 1.3 − 9 |
| M5 | R3 | 48 | 149×31 | 4 | cathode | 0.3 − 5 |
| M5 | R4 | 192 | 149×31 | 4 | wire | 0.1 − 1.5 |

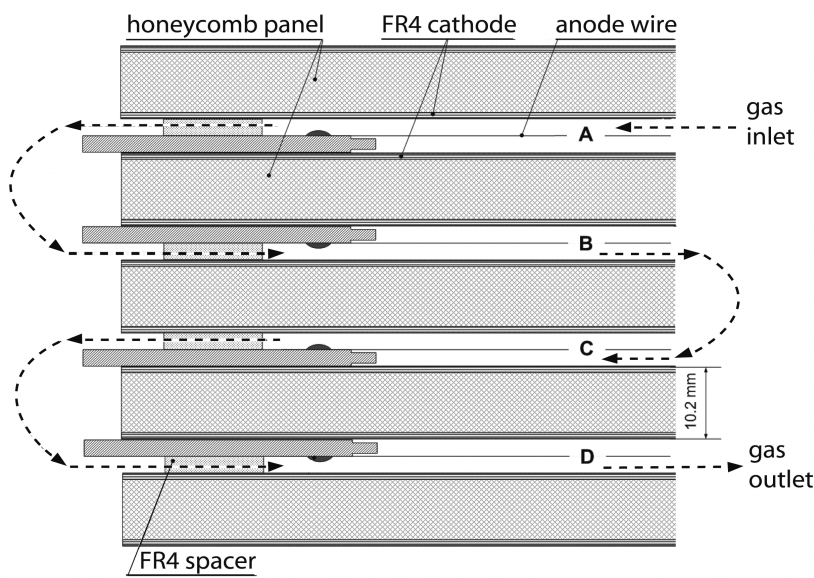

**Figure 2**. Cross section of a MWPC with four gaps indicated by A, B, C and D. The direction of the gas flow is shown by the dashed lines. On the detector, a maximum of 8 MWPCs are connected in series on the same gas line, for a total internal flux from 3 to 16 gas volumes per day, depending on the detector region. It should be noticed that the Muon gas system works in closed loop, recirculating tipically 90% of the gas.

Despite their different dimensions, each chamber has the same internal geometry apart from the number of gaps, as shown in figure 2. Anode planes are centered inside 5 mm gas gaps and are formed by 30 μm diameter gold-plated tungsten wires, with 2 mm spacing. The cathodes are made of FR4 fiberglass plates with two-sided 35 μm thick copper coating. In regions R1−R3 the cathodes have an additional gold coating of about 100nm. Adjacent gaps are separated by panels made of honeycomb or rigid polyurethane foam, which provide precise gap alignment over the whole chamber area. The MWPCs in stations M2−M5 have four gaps, while station M1 is equipped with two-gap MWPCs, giving a total instrumented area of about 1650 m$^2$ (counting each gap separately). The total number of MWPC gaps in the system is 4944. Each of them is powered by an independent HV channel. The corresponding pads in the different gaps of the same chamber are OR-ed in the readout .

The MWPCs are fed with a 40% Ar + 55% CO$_2$ + 5% CF$_4$ gas mixture. This mixture was chosen following studies aiming at optimizing the charge deposit and drift velocity [7]. The presence of CF4 helps the initial cleaning of the electrod surfaces and prevents the formation of Si deposits during MWPC operation. It is also a benefit to suppress ME currents, as it has been verified in the tests of chambers conditioning before installation, as described in chapter 2.1. The gas mixture is supplied into the gaps sequentially, as shown in figure 2. The HV settings, within the detector efficiency plateau, vary between 2.53 and 2.63 kV. The corresponding gas gain ranges between $4.4\times10^4$ and $8.6\times10^4$, as shown in figure 3 [8].

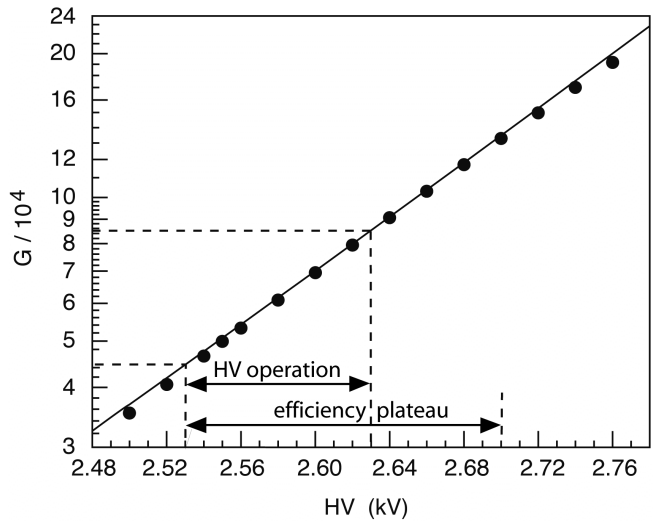

**Figure 3.** Gas gain as a function of HV setting in the LHCb muon MWPCs. The efficiency plateau and the HV operation range are indicated

**2.1 Chamber conditioning before installation**

Before installation on the detector, all MWPCs were conditioned to satisfy severe requirements on the maximum allowed dark current. The conditioning criteria have been established and refined on the basis of the experience acquired in the tests performed during the long phase of design and construction [9]. The chambers must operate steadily for several hours at a positive HV of 2.85 kV with dark current below 10 nA per gap, and at a negative HV of − 2.3 kV with dark current below 150 nA. These requirements were usually satisfied only after a long training phase. We also observed that the chambers of region R3 having gold coated segmented cathodes, required, on average, a longer treatment than the chambers of region R4 having copper cathodes of the same size but not segmented. Due to these features we think that the gaps of region R4 could be cleaned more safely and more efficiently than the ones of region R3, in the phase of chamber construction.

The aim of the conditioning procedure was to remove the residual dust and other pollution left during chamber construction. In particular, operating the detector with negative polarity generates a current of high energy electrons near the cathode which increases the number of active radicals capable of chemical etching of dielectric films. This is of great help in removing possible sources of ME (see appendix A).

Moreover, most of the chambers of the inner regions R1 and R2 passed an additional training phase with positive HV in the range of 2.2 – 2.75 kV, while exposing the chambers to a high gamma ray source at the CERN GIF [10]. The average exposure time per chamber was about 48 hours, with an estimated deposited charge on wire of about 1 mC/cm. On that occasion, the appearance of high self-sustained currents was already observed and the effectiveness of HV training under irradiation to suppress currents caused by ME, when using CF4 in the gas mixture, was demonstrated [11].

## 3. Recovery of chamber gaps affected by HV trips

Over the years the muon system was exposed to very different particle flows in its different areas as shown in the last column of table 1. While no significant ageing effects in terms of reduced gas gain have been observed, even in the most irradiated regions of the system, about 100 MWPC gaps were affected every year from the appearance of high currents. This effect originates from a localized area in the individual gaps of the chambers and results in an increased noise rate and in a trip of the HV supply due to a current exceeding the set threshold. Most of the trips were successfully recovered in situ, during data taking, by means of a HV training with the nominal gas mixture, as discussed in the following.

During the training a relatively high current is maintained, which is a good strategy for recovery from ME (see appendix A). A higher current, in fact, increases the concentration of fluorine radicals, produced by $CF_4$, which react with deposits like silicone and polymers, leading to surface etching by means of the creation of volatile products in the plasma. The high current is maintained independently of the presence of colliding beams.

A typical example of the appearance of a self-sustained current, and the recovery procedure implemented in one of the MWPC gaps of the region M5R3 is shown in figure 4. For this chamber type, operating at a nominal voltage of 2.6 kV, the maximum current value in the presence of colliding beams is ~ 0.6 µA. As shown in the figure 4 (left), the HV channel powering this gap tripped due to the ignition of a high current exceeding the safety threshold set at 30 µA. Following the trip, a specially developed algorithm reduces the HV setting, in steps of 50 V, until the measured current is again below the safety threshold. The operator is alerted as well, who then starts the training procedure.

During the training the HV is automatically decreased, or increased up to a limit of 2.75 kV, to keep the current in a range of high but safe values (30 ± 4 µA in this case). The MWPC is considered recovered when the current at the maximum HV accepted for the training (2.75 kV), drops to the normal value expected at this voltage in presence of colliding beams, and drops to zero in the absence of beams. If the same stable condition is kept for multiple days of operation with beams, the trained gap is put back in normal operation at the nominal voltage. The full recovery cycle for this gap is shown in figure 4 (right). In this particular case, after about two weeks, the self-sustained current definitively dropped down to zero. In more detail, three different cycles of recovery are visible (indicated as A, B and C in figure 4 right). The first cycle was completed in about three days, then a self-sustained current due to ME was initiated again in the presence of colliding beams, and the HV had to be reduced accordingly. The second cycle lasted for about one week, but at the end, the high current appeared again. Only after a third training period of about one week, the gap was fully recovered. The most probable explanation of this complex phenomenology is the subsequent appearance of ME in three different local regions in this particular gas gap. The probability to initiate a high current in the gap under recovery is much higher than under normal conditions, because of the higher voltage applied during training (2.75 kV**),** which provides a gas gain about three times larger than under nominal conditions (2.6 kV). The increased current of positive ions toward the cathode can then generate new zones where the thickness of a dielectric deposit no longer permits the full current flow from the film surface to the underlying cathode, causing an electric field sufficient to generate spontaneous secondary emission of electrons (see appendix A).

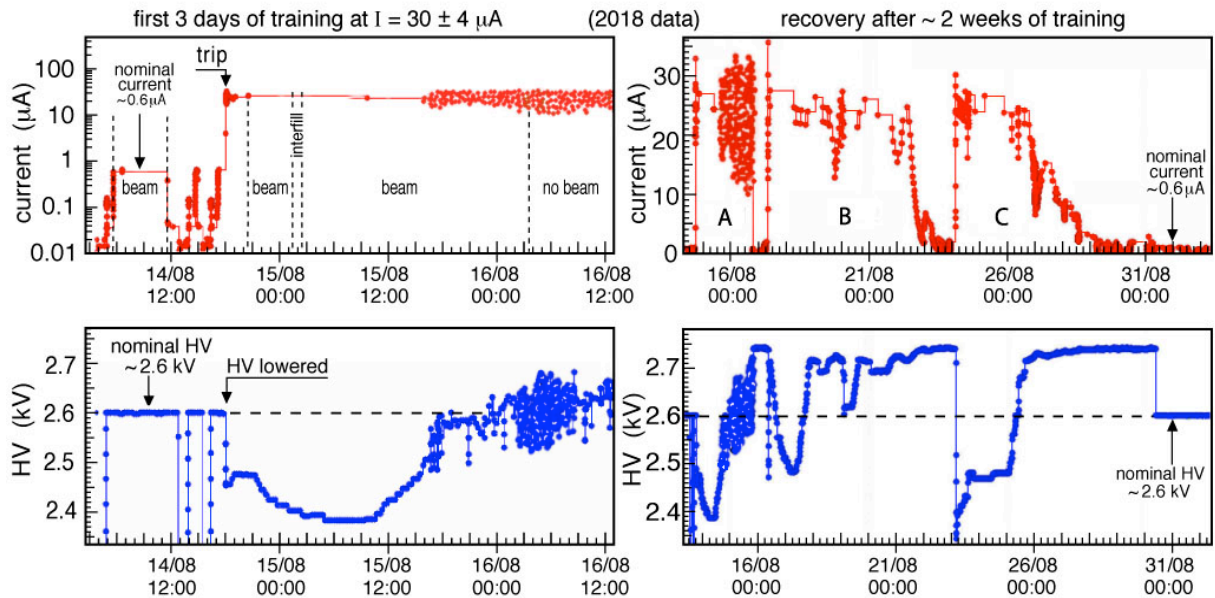

**Figure 4.** A typical example of the appearance of a self-sustained current and of the recovery procedure during normal LHCb operation with beams. The data refer to a gap in region M5R3. The plots on the left show the current (top) and HV setting (bottom) during a period of about 3 days around the first appearance of the HV trip and the subsequent start of HV training. The plots on the right show the current (top) and HV setting (bottom) during the full recovery procedure, which lasted about two weeks. The nominal HV setting for this gap is 2600 V; the average current in presence of colliding beams is about 0.6 μA.

The gap taken as an example in figure 4 was recovered in a relatively short time. The average duration of the training procedure is about two months, but in some cases it requires up to four months for the full recovery. Due to that, at the end of each year of operation, we concluded with up to 25-30 individual gaps under recovery procedure at the same time. It is worth noting though, that during this long training period, the detector efficiency is only marginally affected because the other gaps of the chamber, OR-ed in the readout, are operated at the standard conditions. Moreover in the gaps under training, a sufficiently high voltage is kept for most of the time and most of the readout channels remain efficient, since the high current concerns only a localized area. For these reasons we have never been able to measure any appreciable efficiency reduction during training.

In addition to the training performed during data taking, when access to the detector is possible (mainly during the year-end-technical-stops) all chambers that were affected by high current problems are passed through a conditioning procedure at negative and positive HV polarity. This training aims at reaching the same conditions required for the chambers before installation, as mentioned in the previous section. The same training was also applied preventively to all the chambers of the regions where the greatest number of trips was observed, with the result that approximately 50% of the gaps have undergone this treatment during nine years of operation.

## 4. Statistics of HV trips and training results

Since the start of the LHC in 2010, a total of 375 out of 4944 MWPC gaps, corresponding to 8% of the total, were affected by trips due to ME and treated in situ with the method described above. Only 27 gaps, 0.5% of the total, could not be restored to normal operation. Nine of these developed a short circuit inside the gas gap and the chambers had to be replaced. During the first long shutdown (LS1) of LHC in 2013-2014, four chambers were removed from the experiment, as they contained six gaps that could not be recovered, even after several months of the training procedure. They were recovered using a modified gas mixture with a small addition of oxygen (see Section 5). All of them were installed back in the system and worked well during LHC Run 2.

Table 2 reports the total number of gaps affected by HV trips observed every year of data taking since 2010. For each year the run duration, defined as the number of effective days with colliding beams, the integrated luminosity ($L_{int}$) and the peak luminosity ($L_{peak}$) are also reported. The number of affected gaps is split into gaps tripping for the first time ("new trips") and gaps that had already tripped in the previous years ("recurrent trips").

**Table 2.** Effective run days with colliding beams, integrated luminosity ($L_{int}$) and peak luminosity ($L_{peak}$) per year since 2010. Number of gaps tripping for the first time (new trips) and number of gaps which had already tripped in the previous years (recurrent trips).

| Year | 2010 | 2011 | 2012 | 2015 | 2016 | 2017 | 2018 | Total |
|---|---|---|---|---|---|---|---|---|
| Effective run days | 29 | 56 | 76 | 39 | 86 | 80 | 72 | 438 |
| $L_{int}$ (pb$^{-1}$) | 40 | 1220 | 2210 | 370 | 1910 | 1990 | 2460 | 10200 |
| $L_{peak}$ ($10^{32}$ cm$^{-2}$ s$^{-1}$) | 1.7 | 3.8 | 4.0 | 3.5 | 3.7 | 3.5 | 4.4 | − |
| New trips | 90 | 84 | 69 | 11 | 76 | 18 | 27 | 375 |
| Recurrent trips | 0 | 31 | 67 | 15 | 74 | 32 | 32 | 251 |

The interpretation of the data about HV trips is complex, since there are several factors that are expected to influence the number of HV trips, such as the duration of the run, the detector active area and the local values of particle flux. In addition, the chambers in different regions/stations have slightly different characteristics and different conditioning procedures prior to installation. The trips are thus not randomly distributed, neither in time nor in their position on the detector. Regarding the time distribution, a higher trip frequency was observed when the luminosity was in the ramp-up phase, although the integrated luminosity was rather low in those periods. On the other hand, regarding the distribution of the position of trips on the detector, some MWPC gaps appear to be more susceptible to trips from the beginning, having a higher probability of manifesting high current problems. These critical gaps are not uniformly distributed among the regions and stations. However, no significant correlation with the particle flux is observed, despite the flux strongly differs for the various detector regions.

In figure 5 the cumulative number of MWPC gaps affected at least once by HV trips, normalized to the total number of gaps (4944), is shown as a function of the total number of effective run days integrated between 2010 and 2018. The percentage of gaps affected after nine years of operation is below 8% which reflects the high quality standards maintained during the chamber construction. The 375 gaps affected are distributed over 259, out of the 1368 MWPCs, demonstrating that most of the time only a single gap in a chamber is affected by ME. Figure 5 shows also that the curve is flattening out over the years of system operation.

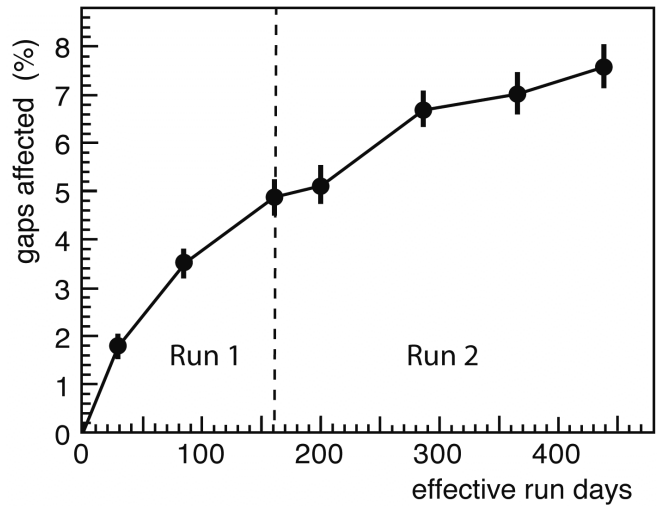

**Figure 5.** Accumulated fractional number of new trips observed in the detector, normalized to the total number of gaps (4944), as a function of the total number of effective run days integrated between 2010 and 2018. The values are evaluated at the end of each year of data taking. Errors are statistical.

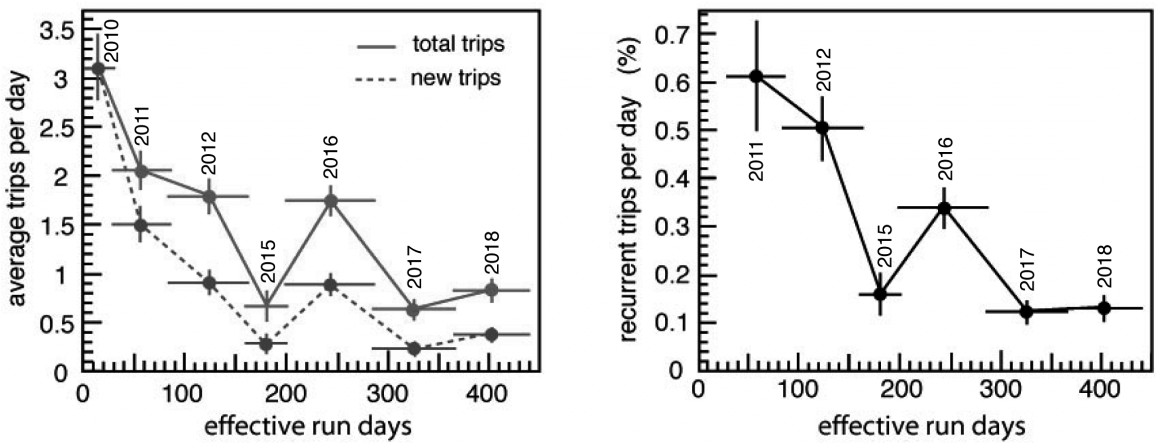

**Figure 6.** Left: Average number of trips per day observed during each year of data taking, as a function of the total number of effective run days integrated between 2010 and 2018; the total number of trips is shown in red (full line), the number of new trips in blue (dotted line). Right: Average number of recurrent trips per day observed during each year of data taking normalized to the total number of gaps already tripped in the past. Points are evaluated starting from 2011. Errors are statistical.

The average number of affected gaps per day observed during each year of data taking is reported in figure 6 (left) for all gaps and for the ones affected for the first time. First of all we notice that the number of gaps with recurrent trips is much higher than expected by a random effect, given the very small percentage of gaps concerned. We also notice a decreasing trend of new trips, suggesting causes originating from the construction phase rather than from ageing or other possible problems occurring during the run. An exception to the decreasing trend is observed in the 2016 run, when we suffered from a sharp increase in the trip frequency, due to a sudden significant increase of the chamber gas gain. The cause could not be fully explained, but seems to be linked to a temporary unwanted change of the gas mixture. We notice also that the two minima of the trip frequency corresponds to the two runs (years 2015 and 2017) in which the peak luminosity (and thus the chamber currents) did not exceed the values reached in previous years.

Last but not least, it can be seen that the frequency of recurrent trips remains about constant even if the total number of gaps that had a trip in the past is steadily increasing with time. This is also shown in figure 6 (right) where the number of recurrent trips per day is normalized to the total number of gaps that have already had a trip in the past. The recurrences observed during the last two years is around 0.1% per day, equivalent to less than 10% probability during the full run. This demonstrates that the adopted training procedure is effective.

As mentioned at the beginning of this section, key features that would be expected to increase the probability for the detector to manifest high current problems are the local particle flux and the detector active area.

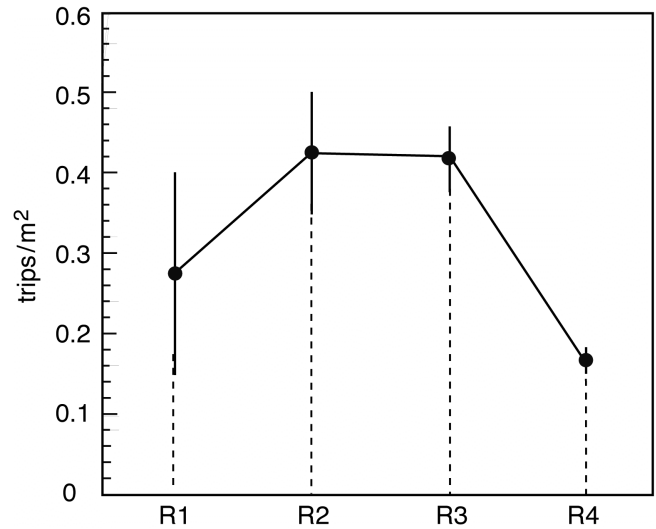

**Figure 7.** Number of HV trips in each region of the detector, normalized to the total instrumented area in the region, with statistical error.

In figure 7 the total number of HV trips per square meter of the detector active area, occurred in 10 years of operation, is shown separately for the four regions exposed to a particle flux per unit area strongly decreasing from R1 to R4. Despite the large statistical errors affecting the values in regions R1 and R2, based on 5 and 33 events respectively, the expected decrease with the radiation dose from region R1 to R3 seems to be contradicted. This is certainly due to the different treatment to which the chambers have been subjected. In particular, as discussed in section 2, most of the chambers of the inner regions R1 and R2 passed before installation through an additional step of the training procedure under irradiation at the GIF, which strongly reduced the trip probability during the following runs at LHC. Conversely the drop of trips in region R4, as compared to region R3 seems to follow the expected behavior, given by the reduced irradiation. In this case however, the observed trend could also be influenced by the better initial cleaning of the R4 gaps as mentioned in chapter 2.1

As a conclusion, the overall picture suggests that most of the HV trips suffered during the detector operation are connected with imperfections existing since the chamber construction, like patches of thin polymer films on the metal cathode surface (see appendix A). The presence of a given thickness of insulation will generate a ME only when the rate of the charge deposited on the cathode will exceed a given threshold following a luminosity increase to values never reached before (or an equivalent increase of the gas gain). It is then evident the primary importance of both an accurate cleaning in the construction phase of the gaps, and the implementation of conditioning procedures, capable of chemical etching of the dielectric films, prior to chamber installation and during shutdown periods before exposition to colliding beams at increasing luminosity.

## 5. Accelerated recovery of MWPCs with addition of oxygen in the gas mixture

Despite the success of the method described above, the long duration of the training procedure introduces complications for its application. This is mostly related to the necessary interruptions of the recovery procedure mainly due to planned stops of the LHC and of the detector. With the aim of reducing the duration of the recovery procedure, a new approach was investigated, consisting of the addition of a small amount of oxygen to the gas mixture during the training procedure. In appendix B a detailed description of the reactions occurring in the gas is reported.

To investigate the effect of the Oxigen, four MWPC chambers were tested during the first long shutdown LS1, in 2013-2014. These chambers were removed from the experimental setup because of the persistence of too high local currents, after multiple failures of the standard recovery procedure in the presence of colliding beams. The localization of the cathode regions affected by ME on the concerned MWPC gaps was performed in the laboratory, with a collimated $^{90}$Sr β-source. Currents a hundred times larger than the ionization current were triggered by the $^{90}$Sr source. As a result of a scan performed on each of the four MWPC gaps of each chamber, six zones affected by ME were identified, as listed in table 3. The zones were randomly distributed on the gap active area.

For each of the above zones, a recovery training as the one described in section 3 was first tried for few hours with the standard gas mixture in order to confirm what was observed on the apparatus. Then a new training was performed with a new gas mixture, adding ~ 2% of oxygen to the nominal one. In all cases the concerned area was kept under irradiation by the $^{90}$Sr β-source. Figure 8 shows the results obtained for one of the gaps. With the standard gas mixture no current decrease is observed after more than 6 hours, consistent with what was obtained in the training with colliding beams. In the presence of oxygen, however, the initial current of 25 μA (initiated at the voltage of 2.6 –2.7 kV) drops sharply until it rises again in correspondence with a voltage increase carried out to maintain a high current level. After a few iterations the current finally drops to zero. The total time required to recover the ME zone in this case was around four hours. Similar results were obtained for all zones affected by ME, as listed in table 3. After the training with oxygen, all of the MWPCs above have been installed back on the apparatus and worked properly, with no recurrent trip during the full Run 2 of the LHC.

Table 3. Recovery time (column 4) measured in 6 zones of different chambers (column 1 and 2) affected by ME, when a 2 % of oxygen was added to the normal gas mixture. In column 2 the gaps are named A, B, C and D according to figure 2.

| Chamber Type | Number of detected ME zones | ME ignition voltage (kV) | Time for recovery (h) |
|---|---|---|---|
| M2R4 | 2 (gap A) | 2.75 | 3 |
| | | 2.8 | 1 |
| M4R4 | 1 (gap D) | 2.7 | 5 |
| M5R4 | 1 (gap A) | 2.6 | 4 |
| | 1 (gap B) | 2.7 | 3 |
| | 1 (gap D) | 2.8 | 2 |

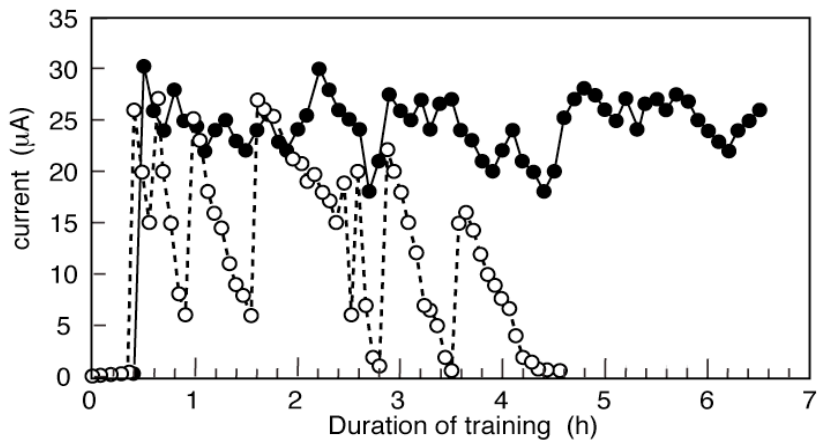

**Figure 8.** Current in the MWPC during the ME recovery training: nominal mixture (full circles) is compared with a mixture containing 2% of oxygen (open circles).

The above tests confirm the hypothesis of an accelerated recovery procedure in presence of oxygen [12], based on plasma chemical etching of silicon and organic compounds by electronegative active radicals and ozone produced in the gas discharge, as discussed in appendix B. Volatile compounds created during the etching are then removed by the gas flux. The amount of oxygen added to the mixture must be small because the electronegative nature of oxygen reduces the electron density in the discharge plasma resulting in the reduction of the charge amplification. By adding 2% of oxygen, the gas gain decreases by a non negligible amount of about 20%, but is still sufficient to support a relatively high current for an effective etching in the recovery process. A small amount of oxygen could be added in the future to the working gas mixture during the year-end-technical-stops, either for targeted recovery interventions, or for conditioning while exposing the chambers to a high intensity radioactive source. Also a permanent use during detector operation of a gas mixture with small amount of Oxygen could be envisaged and fine tuned.

This approach could be particularly useful in view of the LHCb upgrade, which targets a luminosity increase of a factor five in 2021. While it is difficult to predict now the impact of ME on the detector operation at the upgrade conditions, the above result represents a valid strategy to accelerate the chamber recovery and to guarantee an efficient operation into the High-Luminosity-LHC era.

## 6. Conclusions

In nine years of operation in a high radiation environment, the LHCb muon detectors did not show a gain reduction or any other apparent deterioration in performance. However, during this period, about 19% of the 1368 MWPC chambers were affected by high current problems in one of their gas gaps, resulting in HV trips. Most of them are due to Malter-like effects, characterized by the appearance of local self-sustained high currents. The analysis of the trip distribution in time and in the different regions of the detector indicates that ageing due to prolonged irradiation is not the underlying cause and that most of the trips are connected with imperfections existing since the chamber construction.

A method for a non-invasive recovery in-situ has been developed and applied individually to all problematic gaps, consisting of a HV training in the presence of colliding beams and with MWPCs working with the standard gas mixture. The training procedure must normally be continued for a long time (two months on average), before the gaps are recovered and can be restored back to normal operation. Nevertheless, the whole LHCb muon detector could be kept close to 100% efficiency for almost a decade. The redundancy built into the muon detector, using multi-gap MWPC, was crucial to obtain this result.

The recovery procedure developed and tuned over the years has been shown to be effective. Less than 1% of the chambers had to be replaced because of HV trips in 9 years of operation. The percentage of gaps treated with this method in the past and showing repeated high current problems decreased with time and was measured to be about 10% during the last two years of LHC operation.

With the purpose of making the training procedure faster and even more efficient, a method for accelerated recovery has been investigated and successfully tested, using a small (~2%) amount of oxygen in the nominal gas mixture. This can be an important ingredient for the efficient operation of the muon system at increased luminosity.

In general, the non-invasive character of the recovery technique discussed in this paper makes it appealing for many experiments, where the detectors operate with gas mixtures containing $CF_4$.

## Acknowledgments

We express our gratitude to our colleagues in the CERN accelerator departments for the excellent performance of the LHC. We thank the technical and administrative staff at CERN and at the LHCb institutes.

## A. Curing Malter-like effects in MWPC in presence of $CF_4$

The presence of a thin (~1 μm) dielectric polymer film on metal cathode surfaces results in the appearance of a self-sustained secondary emission current, referred to as the Malter effect in the literature [4].

Depending on the value of the positive electric charge deposited on the dielectric film and on the film thickness, the resulting electric field in the dielectric may become sufficient to cause spontaneous secondary emission of electrons from the cathode. This occurs when the formation of the electric charge is not compensated by a leakage current from the film surface to the underlying cathode. The deposit of a thin insulator on the cathode can remain from the MWPC construction process. For example, the chamber panels are produced in a mold, where polyurethane foam is injected in between two printed circuit boards which form the cathode surface. A mold release agent (ACMOIL36-4600) is used, containing 5-10% silicone. This product is suspected to create patches of insulating film on the cathode surface, as numerous studies have shown [13,14].

The presence of $CF_4$ in the working gas mixture of the MWPCs allows for curing the ME via the production of fluorine radicals reacting with silicone and polymers and leading to surface etching by means of the creation of volatile products in the plasma.

In proportional chambers, the most intense formation of free radicals takes place around anode wires where the electric field reaches 20–200 kV/cm. Dissociation of $CF_4$ molecules resulting from the impact of electrons of about 3−5 eV occurs with formation of the following chemically active radicals [15,16]:

$$e^- + CF_4 \rightarrow CF_3^+ + F\bullet + 2e^- \quad (A.1)$$

$$e^- + CF_4 \rightarrow \bullet CF_3 + F\bullet + e^- \quad (A.2)$$

$$e^- + CF_4 \rightarrow \bullet CF_2 + 2F\bullet + e^- \quad (A.3)$$

$\bullet CF_3$, $\bullet CF_2$, $F\bullet$ radicals produced in plasma-chemical reactions (A.1–A.3) efficiently react with different silicon formations. Volatile molecules ($CO_2$, $O_2$ and $SiF_4$) formed in the following etching reactions are easily removed from the detector volume by the gas flow:

$$4F\bullet + Si \rightarrow SiF_4 \uparrow \quad (A.4)$$

$$4F\bullet + SiO_2 \rightarrow SiF_4 \uparrow + O_2 \uparrow \quad (A.5)$$

$$\text{Si} + \bullet\text{CF}_3 + \text{F}\bullet + 2\text{O} \rightarrow \text{SiF}_4\uparrow + \text{CO}_2\uparrow \qquad (A.6)$$

Thus, to recover MWPCs from ME caused by silicone or organic films on the cathode surface, the corresponding depositions can be etched in the presence of $CF_4$. However, in the vicinity of the cathode, which is several millimeters away from the anode wires and the plasma environment, the concentration of fluorine radicals is small. Thus, the recovery procedure often requires relatively long time and sometimes can be inefficient.

In the conditioning procedure at inverse polarity, which can be applied to the chambers during shutdown periods, electrons are accelerated toward the cathode and the presence of high energy electrons increases the concentration of fluorine radicals, capable of chemical etching, near the cathode surface.

## B. Gas compositon for accelerated recovery from Malter effect

Various studies of dry etching processes showed that the etching rate in a $CF_4/O_2$ mixture is significantly higher in comparison to the one in a pure $CF_4$ plasma [16, 17]. Oxygen radicals promote the formation of $\bullet COF_x$, which quickly dissociates in collisions with surrounding electrons and atoms and indirectly increases the number of fluorine radicals in the gas discharge plasma, as shown by the following reactions:

$$\text{O}\bullet + \bullet\text{CF}_3 \rightarrow \bullet\text{COF}_2 + \text{F}\bullet \qquad (B.1)$$

$$\text{O}\bullet + \bullet\text{CF}_2 \rightarrow \bullet\text{COF} + \text{F}\bullet \qquad (B.2)$$

$$e^- + \text{COF}_2 \rightarrow \bullet\text{COF} + \text{F}\bullet + e^- \qquad (B.3)$$

$$\text{O}\bullet + \bullet\text{COF} \rightarrow \text{CO}_2\uparrow + \text{F}\bullet \qquad (B.4)$$

Moreover, oxygen itself plays a significant role in polymer film dry etching [18]. Kinetics of chemical reactions in MWPC gas discharges may significantly differ from reaction rates in industrial reactors due to the different plasma nature, gas pressure and electric field configuration. In MWPCs molecules and radicals have significantly smaller mean free paths between electron collisions, nevertheless the average electron energy (5-10 eV) is quite similar to the one in reactors [19]. This makes it possible to use dry etching chemical models in qualitative predictions of the chemical processes in MWPC. Relevant reactions of oxygen dissociation and excitation by impact of electrons in this energy range are given below:

$$e^- + \text{O}_2 \rightarrow \text{O}^- + \text{O}^{\bullet\bullet} \qquad (B.5)$$

$$e^- + \text{O}_2 \rightarrow {}^*\text{O}_2 + e^- \qquad (B.6)$$

Both the atomic oxygen, $O^{\bullet\bullet}$, and the excited molecular oxygen, $^*O_2$, are chemically aggressive. Atomic oxygen $O^{\bullet\bullet}$, interacts with $O_2$ molecules forming ozone, $O_3$, which can participate in the processes of plasma chemical etching on the cathode surface or recombine with atomic oxygen into $O_2$ molecules.

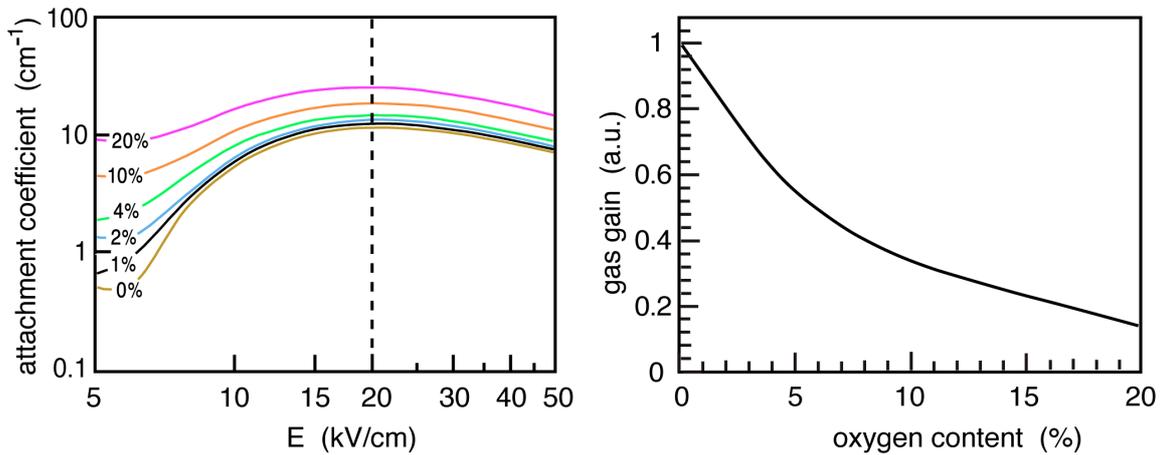

**Figure 9.** Left: electron attachment coefficient as a function of the electric field strength. Right: the gas gain as a function of the oxygen content.

However, $O_2$ content in MWPC working gas mixture should be strongly limited. In fact, due to its high electron attachment coefficient, the oxygen reduces the electron density in the discharge plasma. This results in the reduction of the charge amplification. Thus to keep the gas gain at the level sufficient for the recovery process, the oxygen content should be optimized.

To find the optimal amount of oxygen to be added to the working gas mixture, simulation studies for the muon detector MWPCs have been performed using the GARFIELD software package [20]. The results of the simulation are shown in figure 9. The electron attachment coefficient as a function of the electric field strength is shown in the left plot. The right plot shows the gas gain dependence on the oxygen content, where the latter is added to the standard 40% Ar + 55% CO2 + 5% $CF_4$ gas mixture.

As a result, for 1-4% $O_2$ content the electron attachment coefficient increases substantially only in the drift region, especially near the cathode surface, where the electric field strength is about 6 kV/cm. Conversely, when oxygen content exceeds 10%, a noticeable electron attachment occurs throughout the whole drift path and the avalanche region. As a result, the MWPC gas gain at the operation voltage would drop by more than 60%, too much to be compensated by an increase of HV sufficient to support an effective etching process. In the test of accelerated recovery of the MWPCs a percentage as small as ~2% of oxygen was added to the nominal gas mixture and satisfactory results were obtained.